\documentclass[a4paper,fleqn]{cas-dc}

\usepackage[numbers]{natbib}

\def\tsc#1{\csdef{#1}{\textsc{\lowercase{#1}}\xspace}}
\tsc{WGM}
\tsc{QE}
\tsc{EP}
\tsc{PMS}
\tsc{BEC}
\tsc{DE}

\def\be{\begin{equation}}
	\def\ee{\end{equation}}
\newcommand{\bel}[1]{\begin{eqnarray}\label{#1}}
	\newcommand{\eel}{\end{eqnarray}}
\def\barr{\begin{array}}
	\def\earr{\end{array}}
\def\beq{\begin{eqnarray}}
	\def\eeq{\end{eqnarray}}
\def\bfig{\begin{figure}}
	\def\efig{\end{figure}}

\newcommand{\nn}{\nonumber}
\newcommand{\f}[2]{\frac{#1}{#2}}
\newcommand{\onehalf}{{\nicefrac{1}{2}}}

\newcommand{\p}{\partial}

\usepackage{amssymb}

\usepackage{graphicx}
\usepackage{amsmath}
\usepackage{subfigure}
\usepackage{bbold}
\usepackage{yfonts}
\usepackage{placeins}
\usepackage{bm} 
\usepackage{nicefrac}
\usepackage{slashed}
\usepackage{lipsum}

\newcommand{\bea}{\begin{eqnarray}}
\newcommand{\eea}{\end{eqnarray}}

\newcommand{\EQ}[1]{Eq.~(\ref{#1})}

\def\spin{\,\mathfrak{s}}

\newcommand{\pv}{{\boldsymbol p}}

\def\kv{{\boldsymbol {k}}}

\def\oabU{\omega^{\alpha\beta}}
\def\omnLD{{\tilde \omega}_{\mu\nu}}

\def\epsUabgd{\epsilon^{\alpha \beta \gamma \delta}}

\def\epsLmnab{\epsilon_{\mu\nu\alpha\beta}}

\def\SmnU{{\Sigma}^{\mu\nu}}

\def\S0iU{{\Sigma}^{0i}}

\def\n0{n_{(0)}}
\def\e0{\varepsilon_{(0)}}
\def\P0{P_{(0)}}

\def\Wxk{{\cal W}(x,k)}

\def\Fpxk{{\cal F}^{+}(x,k)}
\def\Fmxk{{\cal F}^{-}(x,k)}

\def\Fxk{{\cal F}(x,k)}

\def\Pxk{{\cal P}(x,k)}

\begin{document}
\let\WriteBookmarks\relax
\def\floatpagepagefraction{1}
\def\textpagefraction{.001}
\shorttitle{Relativistic dissipative spin dynamics in the relaxation time approximation}
\shortauthors{S. Bhadury et~al.}

\title [mode = title]{Relativistic dissipative spin dynamics in the relaxation time approximation}                      



\author[1]{Samapan Bhadury}[
orcid=0000-0002-8693-4482
]
\ead{samapan.bhadury@niser.ac.in}


\address[1]{School of Physical Sciences, National Institute of Science Education and Research, HBNI, Jatni-752050, India}

\author[2]{Wojciech Florkowski}[
                       orcid=0000-0002-9215-0238
                        ]
\ead{wojciech.florkowski@uj.edu.pl}


\address[2]{Institute of Theoretical Physics, Jagiellonian University, PL-30-348 Krakow, Poland}

\author[1]{Amaresh Jaiswal}[orcid=0000-0001-5692-9167]


\ead{a.jaiswal@niser.ac.in}

\author[3]{Avdhesh Kumar}[orcid=0000-0003-1335-8880]
\cormark[1]
\ead{avdhesh.kumar@ifj.edu.pl}

\author[3]{Radoslaw Ryblewski}[ orcid=0000-0003-3094-7863
                        %
   ]
\ead{radoslaw.ryblewski@ifj.edu.pl}


\address[3]{Institute of Nuclear Physics Polish Academy of Sciences, PL-31-342 Krakow, Poland}



\cortext[cor1]{Corresponding author}


\begin{abstract}
The concept of the Wigner function is used to  construct a semi-classical kinetic theory describing the evolution of the axial current phase-space density of spin-$\onehalf$ particles in the relaxation time approximation. The resulting approach can be used to study spin polarization effects in relativistic matter, in particular, in heavy-ion collisions. An expression for the axial current based on the classical treatment of spin is also introduced and we show that it is consistent with earlier calculations using Wigner functions. Finally, we derive non-equilibrium corrections to the spin tensor, which are used to define, for the first time, the structure of spin transport coefficients in relativistic matter.
\end{abstract}



\maketitle

\section{Introduction}

The spin polarization of various particles ($\Lambda$, $K^*$, $\phi$) produced in relativistic heavy-ion collisions has been recently observed by the STAR experiment at the Brookhaven National Laboratory (BNL) \cite{STAR:2017ckg,Adam:2018ivw,Acharya:2019vpe}. In the case of the $\Lambda$ hyperons, a quite substantial global polarization (of about 10\%, due to the spin-orbit coupling) was theoretically forseen in Refs. \cite{Liang:2004ph,Betz:2007kg,Voloshin:2004ha}. However, a smaller polarization (of about 1\%, due to equilibration of spin degrees of freedom)  was proposed later in Refs. \cite{Becattini:2007nd,Becattini:2007sr,Becattini:2013fla,Becattini:2013vja} and such an effect was eventually observed by STAR \cite{STAR:2017ckg,Adam:2018ivw}. Being the first experimental observation of a non-zero spin polarization in heavy-ion collisions, it has been commonly interpreted as one of the greatest discoveries in physics in 2017~\cite{STAR:2017ckg}.  

While the global spin polarization of the $\Lambda$ (and ${\bar\Lambda}$) hyperons can be explained by the assumption that the spin polarization is directly expressed by the so-called thermal vorticity \cite{Becattini:2007nd,Becattini:2007sr,Becattini:2013fla,Becattini:2013vja,Li:2017slc,Sun:2017xhx}, other features of the data lack convincing theoretical explanations \cite{Niida:2018hfw,Adam:2019srw}. The interpretation problems appear also in the case of the $K^*$ and $\phi$ mesons~\cite{Lisa:2019}. This situation has triggered many theoretical studies, for  example, see \cite{Becattini:2018duy,Hattori:2019lfp,Wu:2019eyi,Sheng:2019kmk,Becattini:2019ntv,Liu:2019krs,Yang:2020hri,Gallegos:2020otk}.

The problems described above suggest that the spin effects in heavy-ion collisions can be independent of the thermal vorticity and governed by other type of dynamics. In Ref. \cite{Florkowski:2017ruc}, a hydrodynamic framework for particles with spin $\onehalf$ was proposed, where the spin polarization is described by the spin polarization tensor $\omega_{\mu\nu}(x)$, whose dynamics follows from the conservation law for the angular momentum --- similarly to the evolution of the local temperature $T(x)$ and the hydrodynamic flow vector $u^\mu(x)$ that follow from the conservation laws for the energy and linear momentum. Recently, this framework (with the updated forms of the energy-momentum and spin tensors, which have been related to the underlying Lagrangian of the Dirac field) has been derived from the kinetic theory \cite{Florkowski:2018ahw} (for a recent review see \cite{Florkowski:2018fap}).

In this work, we go beyond the perfect-fluid setup constructed in \cite{Florkowski:2017ruc}. We first introduce semi-classical kinetic equations describing the evolution of the Wigner function of massive spin-$\onehalf$ particles in the relaxation-time approximation (RTA) \cite{PhysRev.94.511,Baym:1984np}. Then, we switch to the framework based on the classical description of spin. It has been shown in \cite{Florkowski:2018fap} that the classical approach is advantageous as it is not restricted to the case of small spin polarization. At the same time, it reduces to the Wigner-function approach if polarization is small. Moreover, it can be used to determine the structure of dissipative spin corrections in a completely analogous way to that known from the standard RTA. 

Although the kinetic description of matter based on the RTA method may seem to be oversimplified, in the last years this method has turned out to be a very useful tool to address numerous physics problems, often of fundamental importance like testing applicability of hydrodynamics~\cite{Florkowski:2013lya,Denicol:2014xca} or early thermalization and/or hydrodynamization puzzles \cite{Heller:2015dha,Heller:2016rtz,Romatschke:2017vte}. Therefore, we propose here a version of the RTA, for particles with spin, which can play a very similar role in extended studies involving spin as an additional macroscopic degree of freedom. 

In order to obtain a clear physics picture, our considerations are restricted to hydrodynamics and kinetic theory of spin-$\onehalf$, massive particles being on the mass shell. Their polarization is described by the axial current density that can be directly used to define the spin density matrix~\cite{Florkowski:2017dyn,Leader:2001gr}. For the metric tensor, Levi-Civita symbol, and the scalar product, we use the following notation and conventions:  $g_{\mu\nu} =  \hbox{diag}(+1,-1,-1,-1)$, $\epsilon^{0123} = -\epsilon_{0123}~=~1$, $a \cdot b = g_{\mu \nu} a^\mu b^\nu = a^0 b^0 - \bf{a}\cdot\bf{b}$. Throughout the text we use natural units with $c = \hbar = k_B =1$.

\section{Kinetic equations for scalar and axial-vector components of the Wigner function}

The studies of relativistic plasma for particles with spin~$\onehalf$ commonly use the Wigner function $\Wxk$ and its Clifford-algebra decomposition~\cite{Vasak:1987um,Zhuang:1995pd,Elze:1986hq,Elze:1986qd,Florkowski:1995ei, Weickgenannt:2019dks}, 
\bea
\Wxk &=& \f{1}{4} \left[ \Fxk + i \gamma_5 \Pxk + \gamma^\mu {\cal V}_{\mu}(x,k) \right. \nn \\
&& \left.  + \gamma_5 \gamma^\mu {\cal A}_{\mu}(x,k)
+ \SmnU {\cal S}_{\mu \nu}(x,k) \right].
\label{eq:wig_expansion}
\eea
Here $x$ is the space-time coordinate and $k^\mu = (k^0, \kv)$ denotes the particle momentum. The coefficient functions appearing on the right-hand side of Eq.~(\ref{eq:wig_expansion}) are sums of the particle and antiparticle contributions, for example, $\Fxk = \Fpxk + \Fmxk$ and ${\cal A}_{\mu}(x,k) = {\cal A}^+_{\mu}(x,k)+{\cal A}_{\mu}^-(x,k)$. We use the Dirac representation for gamma matrices with $\SmnU = (\nicefrac{i}{4}) [\gamma^\mu,\gamma^\nu]$ being the Dirac spin operator.

From the leading and next-to-leading orders of the semi-classical expansion of $\Wxk$ in powers of $\hbar$, one obtains two independent kinetic equations, for the scalar and axial-vector components~\cite{Vasak:1987um,Zhuang:1995pd,Elze:1986hq,Elze:1986qd,Florkowski:1995ei},
\bel{eq:kineqFC1}
k^\mu \p_\mu {\cal F}(x,k) = C_{\cal F},
\eel
\bel{eq:kineqAC1}
k^\mu \p_\mu \, {\cal A}^\nu(x,k) = C^\nu_{\cal A}, 
\quad k_\nu \,{\cal A}^\nu(x,k) = k_\nu C^\nu_{\cal A} = 0.
\eel
Here, we have neglected the effects of the mean fields (which are widely discussed in the literature) and schematically included complicated effects of collisions by adding collision terms on the right-hand sides of Eqs.~(\ref{eq:kineqFC1}) and (\ref{eq:kineqAC1}).
~\footnote{We stress that the coefficient functions ${\cal F}$, ${\cal A}^\nu$, as well as the collision terms $C_{\cal F}$ and $C^\nu_{\cal A}$ are all of the same (leading) order in $\hbar$. The functions ${\cal F}$, ${\cal A}^\nu$ are also the only two independent coefficients of the Wigner function (\ref{eq:wig_expansion}) within this approximation, hence, Eqs. (\ref{eq:kineqFC1}) and (\ref{eq:kineqAC1}) fully determine the evolution of the Wigner function provided the form of the collision terms is defined. }

If the collision terms vanish, Eqs.~(\ref{eq:kineqFC1}) and (\ref{eq:kineqAC1}) describe free streaming of particles. There are also two other cases that can be analyzed with the help of those equations, namely, the global and local thermodynamic equilibrium. This requires, however, the knowledge of the Wigner function in a local thermodynamic equilibrium which can be generally expressed in terms of the scalar, ${\cal F}_{\rm eq}$, and the axial-vector components, ${\cal A}^\mu_{{\rm eq}}$ \cite{Florkowski:2018ahw,  Florkowski:2018fap, Florkowski:2019gio}. Recently, several works have used this concept to express the equilibrium Wigner function as a phase space integral \cite{Becattini:2013fla,Florkowski:2018ahw, Florkowski:2018fap,Prokhorov:2017atp}
\bea
{\cal F}^\pm_{\rm eq} &=& 2 m\,\int dP\,{f}^{\pm}_{\rm eq}(x,p)\,\delta^{(4)} (k\mp p),  \label{eq:FEeqpm} \\
{\cal A}^\pm_{{\rm eq}, \mu} &=& -  \, \int dP\,\,  \omnLD(x) \,p^{\nu} \, {f}^{\pm}_{\rm eq}(x,p)\,\delta^{(4)}(k\mp p)\, ,\nn\\
\label{eq:AEeqpm}
\eea
where the local equilibrium distributions in the classical (Boltzmann) approximation are given by the expression
\begin{eqnarray}
{f}^{\pm}_{\rm eq}(x,p) = \exp\left[-\beta(x) \cdot p \pm \xi(x) \right].
\label{eq:fpm1}
\end{eqnarray} 
Here $\beta^\mu = u^\mu/T$, where $u^\mu$ is the hydrodynamic flow vector and $T$ is the local temperature, $\xi = \mu/T$ is the ratio of the chemical potential and temperature, $p^\mu = (E_p, \pv)$, and $dP =  d^3p/E_p (2 \pi )^3$, with $E_p = \sqrt{m^2 + \pv^2}$ denoting the on-mass-shell particle energy. The quantity $\omega_{\mu\nu}$ ($\omnLD = \f{1}{2} \epsLmnab  \oabU$) is the spin polarization tensor (dual spin polarization tensor). The former can be interpreted as the ratio of the spin chemical potential $\Omega_{\mu\nu}$ and the local temperature $T$, namely $\omega_{\mu\nu} = \Omega_{\mu\nu}/T$~\cite{Becattini:2018duy}. 
We note that $\omega_{\mu\nu}$ as well as $\omnLD$ are dimensionless quantities in our approach that play a similar role as $\xi$. The formalism presented here is valid for small values of $\omega_{\mu\nu}$ (i.e., for $\omega_{\mu\nu} < 1$), for a discussion of this point see Sec. 6.6 of Ref.~\cite{Florkowski:2018fap}.

\section{RTA for the scalar component}

We start our considerations with the scalar component that has been analyzed in many papers and its treatment is very well established now.
The RTA collision term in the equation for the scalar coefficient takes the following form
\beq 
C_{\cal F} = k\cdot u \,\,
\frac{{\cal F}_{\rm eq}(x,k) - {\cal F}(x,k)}{\tau_{\rm eq}},
\label{eq:kineqFC1m}
\eeq
where $\tau_{\rm eq}$ is the relaxation time.
With the definition
\bea
{\cal F}^\pm(x,k) &=& 2 m\,\int dP\, \,f^\pm(x,p)\,\,\delta^{(4)} (k\mp p)  \label{eq:FEpm} 
\eeq
one finds
\beq
p^\mu \p_\mu {f}^{\pm}(x,p) &=& p \cdot u \, \frac{
	f_{\rm eq}^{\pm}(x,p)-f^{\pm}(x,p)}{\tau_{\rm eq}},
\label{eq:fpm}
\eeq 
which is the Anderson-Witting model for the RTA \cite{anderson1974relativistic}.

In order to switch from the microscopic kinetic theory to the effective hydrodynamic description one takes the moments of \EQ{eq:kineqFC1}. The zeroth and the first moments correspond then to the conservation laws for charge, energy, and linear momentum 
\beq 
\p_\mu N^\mu = 0, \quad \p_\mu T^{\mu\nu} =0.
\label{eq:conlaw}
\eeq 
Here $N^\mu = \int dP \,p^\mu \left(f^+ - f^-\right)$ and, similarly,\\ $T^{\mu\nu} = \int dP \,p^\mu p^\nu \left(f^+ + f^-\right)$. The necessary condition for the conservation laws (\ref{eq:conlaw}) to be valid is that the appropriate moments of the collision term vanish. This leads to so-called Landau matching conditions: 
\beq 
u_\mu N^\mu_{\rm eq} 
= u_\mu N^\mu,  \quad
u_\mu T^{\mu\nu}_{\rm eq}= u_\mu T^{\mu \nu}. 
\label{eq:LNT}
\eeq
These relations are used to determine the parameters appearing in the equilibrium distributions, namely $T$, $u^\mu$, and $\mu$.

\section{RTA for the axial component}

 In this letter, we propose a natural generalization of the RTA approach that is applicable for the axial component,
 \beq 
 C_{\cal A}^\nu  = k\cdot u \,\,
 \frac{{\cal A}^\nu_{\rm eq}(x,k) - {\cal A}^\nu(x,k)}{\tau_{\rm eq}}.
 \label{eq:kineqAC1m}
 \eeq
 With the definition
 \bea
 {\cal A}^\mu_\pm(x,k) &=& 2m \,\int dP\, \,a^\mu_\pm(x,p)\,\,\delta^{(4)} (k\mp p)  
 \label{eq:AEpm} 
 \eeq
 one finds
 \beq
 p^\mu \p_\mu {a}^\nu_{\pm}(x,p) &=& p \cdot u  \,\, \frac{
 	a^\nu_{\pm \rm eq}(x,p)-a^\nu_{\pm}(x,p)}{\tau _{\rm eq}},
 \label{eq:apm}
 \eeq 
 where the local equilibrium distributions are defined as 
 \beq 
 {a}^\nu_{\pm \rm eq}(x,p) = -\frac{1}{2m} {\tilde \omega}^{\nu \mu}(x) p_\mu {f}^{\pm}_{\rm eq}(x,p). 
 \label{eq:anueq}
 \eeq 
 
 The moments of the scalar equation (\ref{eq:kineqFC1}) naturally lead to the conservation laws for charge, energy, and linear momentum. In the case of the axial equation (\ref{eq:kineqAC1}), the situation is less obvious. Nevertheless, if the collisions are absent and no mean fields are present, then the axial equation describes free-streaming of spin degrees of freedom --- the spin polarization tensor should be constant in this case. 
 
 This fact was known long time ago to be in contrast with the behavior of the canonical spin tensor~\cite{Fradkin:1961,Hilgevoord:1962,DeGroot:1980dk}. The latter is obtained by the Noether theorem and is known to be not conserved even for a free Dirac field. This difficulty was overcome by switching from the canonical forms of the energy-momentum and spin tensors to the GLW expressions (GLW stands here for de~Groot, van~Leeuwen, and van~Weert of Ref.~\cite{DeGroot:1980dk}). 
 
 In our case, i.e., for particles being on the mass shell (note the Dirac delta functions in Eqs.~(\ref{eq:FEeqpm}) and (\ref{eq:AEeqpm})), the connection between the axial current and the GLW spin tensor takes a particularly simple form
 \bea
 S^{\lambda, \mu \nu}(x) = \frac{1}{2}
 \varepsilon^{\alpha \beta \mu \nu} \!\!\! \int d^4k   \, \frac{k^\lambda k_\beta }{m^2}
 {\cal A}_\alpha(x,k).
 \eea
 Multiplying Eq.~(\ref{eq:kineqAC1}) by $\varepsilon^{\nu \beta \gamma \delta} k_\beta/m^2$ and integrating over $k$, we find
 \bea
 \partial_\mu S^{\mu, \gamma \delta} = u_\mu \frac{S^{\mu, \gamma \delta}_{\rm eq}-S^{\mu, \gamma \delta}}{\tau_{\rm eq}}.
 \label{eq:LCspin}
 \eea
 Hence, the requirement of vanishing divergence of the GLW spin tensor leads to a constraint
 \bea 
 u_\mu 
 S^{\mu, \gamma \delta}_{\rm eq} = u_\mu  S^{\mu, \gamma \delta}. \label{eq:LS}
 \eea 
 This formula represents an additional, Landau-type, matching condition for the equilibrium distribution function. It allows to determine six independent components of the spin polarization tensor $\omega_{\mu\nu}$ appearing in Eq.~(\ref{eq:anueq}).
 
 In view of the above discussion it seems natural to consider the conservation of the GLW spin tensor as a parallel condition to the conservation of charge, energy, and linear momentum.  The canonical energy-momentum tensor is asymmetric, and differs from the GLW form by quantum corrections. The latter give rise to mixing between orbital and spin components  of the canonical angular momentum. This phenomenon, known as the spin-orbit coupling, is absent in the GLW pseudo gauge provided the collision term is local and the mean fields are neglected~\cite{Hess:1966}.

\section{Approach with classical spin}

Equations (\ref{eq:fpm}) and (\ref{eq:apm}), together with the Landau matching conditions (\ref{eq:LNT}) and (\ref{eq:LS}), form a consistent system of kinetic equations that allows for determination of the space-time evolution of the phase-space distribution as well as spin densities. They are, however, valid only in the case of small polarization tensor. In order to overcome this restriction, one can switch to a classical description of spin degrees of freedom. We have shown in \cite{Florkowski:2018fap} that the results obtained with the Wigner function for small $\omega$ can be exactly reproduced in the framework with the classical treatment of spin~\cite{Mathisson:1937zz}. An advantage of the classical treatment is, however, that it can be applied to systems with arbitrary polarization.

In the classical approach, we introduce the internal angular momentum of a particle, $s^{\mu\nu}$, and the spin vector, $s^\mu$, connected by the relation $s^{\alpha} = (1/2m) \epsUabgd p_\beta s_{\gamma \delta}$. Accordingly, one introduces a classical distribution function $f^\pm_s(x,p,s)$ in an extended phase space that besides space-time coordinates and momenta includes the spin vector. The appropriate normalization is
\begin{eqnarray}
\int dS \, f^\pm(x,p,s) &=& f^\pm(x,p), \\
\int dS s_\mu f^\pm(x,p,s) &=& a^\pm_\mu(x,p),
\label{a}
\end{eqnarray}
where $dS = (m/\pi \spin) \,  d^4s \, \delta(s \cdot s + \spin^2) \, \delta(p \cdot s)$ with the length of the spin vector defined by the eigenvalue of the Casimir operator, $\spin^2 = \f{1}{2} \left( 1+ \f{1}{2}  \right) = \f{3}{4}$. One can easily check that Eqs.~(\ref{eq:fpm}) and (\ref{eq:apm}) can be obtained as the zeroth and first moments in the spin space of the classical RTA equation
\begin{equation}\label{RTA_spin}
p^\mu \partial_\mu f^\pm(x,p,s) = p \cdot u
\, \frac{f^\pm_{\rm eq}(x,p,s)-f^\pm(x,p,s)}{\tau_{\rm eq}},
\end{equation}
where the equilibrium, spin-dependent function reads
\begin{equation}
f^\pm_{\rm eq}(x,p,s) =  f_{\rm eq}^{\pm}(x,p)\exp\left( \frac{1}{2} \omega_{\mu\nu} s^{\mu\nu} \right).
\label{eq:feqxps}
\end{equation}
Moreover, one can check for small values of $\omega$ that the formula (\ref{eq:feqxps}), when used in Eq.~(\ref{a}), yields Eq.~(\ref{eq:anueq}). Consequently, one can use Eq.~(\ref{eq:feqxps}) in Eq.~(\ref{RTA_spin}), together with the Landau matching conditions (\ref{eq:LNT}) and (\ref{eq:LS}), where 
\begin{eqnarray}
N^\mu &=& \int dP dS \, p^\mu \left(f^+_s - f^-_s \right), \\
T^{\mu\nu} &=& \int dP dS \, p^\mu p^\nu \left(f^+_s + f^-_s \right), \\
S^{\lambda,\mu\nu} &=& \int dP dS \, p^\lambda s^{\mu\nu} \left(f^+_s + f^-_s \right).
\label{eq:classNTS}
\end{eqnarray}
In the above equation $f^\pm_s=f^\pm(x,p,s)$.

\section{Dissipative corrections}

It is straightforward to search for solutions of the transport equation \eqref{RTA_spin} in a series form  $f^\pm(x,p,s) = f^\pm_{\rm eq}(x,p,s) + \delta f^\pm(x,p,s) + \ldots  $, which yields 
\begin{equation}
\delta f^\pm(x,p,s) = - \frac{\tau_{\rm eq}}{p \cdot u} \,\, p^\mu \partial_\mu     f^\pm_{\rm eq}(x,p,s) .
\end{equation}
One can define dissipative corrections to the conserved current $\delta N^\mu \equiv N^\mu - N^\mu_{\rm eq}$, energy-momentum tensor $\delta  T^{\mu\nu} \equiv  T^{\mu\nu} -  T^{\mu\nu}_{\rm eq}$, and the spin tensor $\delta S^{\lambda,\mu\nu} \equiv S^{\lambda,\mu\nu} - S^{\lambda,\mu\nu}_{\rm eq}$, in terms of the moments of $\delta f^\pm_s(x,p,s)$ used in Eq.~(\ref{eq:classNTS}).

After straightforward but quite lengthy calculations, one obtains the following expressions valid in the case of small polarization:
\begin{eqnarray}
\delta N^\mu &=& \tau_{\rm eq}\, \beta_n(\nabla^\mu \xi), \label{deltaN}\\
\delta T^{\mu\nu} &=& \tau_{\rm eq} \left( -\beta_\Pi\,\Delta^{\mu\nu}\,\theta + 2\beta_\pi\, \sigma^{\mu\nu} \right), \label{deltaT}\\
\delta S^{\lambda,\mu\nu} &=& \tau_{\rm eq} \Big[ B^{\lambda,\mu\nu}_{\Pi}\, \theta + B^{\kappa\lambda,\mu\nu}_{n}\, (\nabla_\kappa \xi) + B^{\alpha\kappa\lambda,\mu\nu}_{\pi}\, \sigma_{\alpha\kappa}\nonumber\\
&& \qquad + B^{\kappa \beta \alpha \lambda ,\mu \nu }_{\Sigma }\, (\nabla_\kappa \omega_{\beta\alpha}) \Big]. 
\label{deltaS}
\end{eqnarray}
Here $\Delta^{\mu\nu} = g^{\mu\nu}-u^\mu u^\nu$, $\nabla^\mu = \Delta^{\mu\nu} \partial_\nu$, $\theta$ is the expansion scalar, and $\sigma_{\mu\nu}$ is the shear flow tensor. Different  coefficients appearing on the right-hand side of Eq.~(\ref{deltaS}) have tensor structures expressed in terms of equilibrium tensor quantities $u^\mu,~\Delta^{\mu\nu}$ and $\omega^{\mu\nu}$.  In particular, the structure of the new spin coefficients $B^{\mu_1\mu_2....}_X$ appearing in Eq.~ (\ref{deltaS}) is as follows: 
\begin{eqnarray}
B^{\lambda,\mu\nu}_{\Pi} \hspace{-0.3cm}&=&\hspace{-0.3cm}
B_{\Pi }^{(1)}u^{[\mu }\omega ^{\nu ]\lambda }\!+\!B_{\Pi }^{(2)}u^{\lambda }u^{\alpha }u^{[\mu }\omega ^{\nu ]}{}_{\alpha }\!+\!B_{\Pi }^{(3)}
\Delta ^{\lambda [\mu }u_{\alpha }\omega ^{\nu ]\alpha },\nn\\
B_{\pi}^{\alpha\kappa\lambda ,\mu \nu }\hspace{-0.3cm}&=&\hspace{-0.3cm} B_{\pi }^{(1)}\Delta^{[\mu \kappa }\Delta^{\lambda \alpha }u_{\gamma }\omega ^{\nu ]\gamma }+B_{\pi }^{(2)}\Delta^{\lambda \alpha }u^{[\mu }\omega ^{\nu ]\kappa }\nn\\&&+B_{\pi }^{(3)}u^{[\mu }\Delta^{\nu ]\alpha} \Delta^{\lambda}_{\gamma}\omega ^{\gamma \kappa } + B_{\pi }^{(4)}\Delta ^{\lambda [\mu }\omega ^{\rho \kappa } u_{\rho }\Delta^{\nu ]\alpha },\nn\\ 
\label{eq:betapi}
B_n^{\kappa\lambda,}{}^{\mu \nu }&=&B_n^{(1)}\Delta ^{\lambda \kappa } \omega ^{\mu \nu }+B_n^{(2)}\Delta ^{\lambda \kappa }u^{\alpha }u^{[\mu }\omega ^{\nu ]}{}_{\alpha }\nn\\&&+B_n^{(3)}\Delta ^{\lambda \alpha }\Delta ^{[\mu \kappa }\omega ^{\nu ]}{}_{\alpha }+B_n^{(4)}u^{[\mu }\Delta ^{\nu ]\kappa }u^{\rho }\omega ^{\lambda }{}_{\rho }\nn\\&&+B_n^{(5)}\Delta ^{\lambda [\mu }\omega ^{\nu ]\kappa }+B_n^{(6)}\Delta ^{\lambda [\mu }u^{\nu ]}u_{\alpha }\omega ^{\alpha \kappa},\nn\\
\label{eq:betan}
B_{\Sigma }^{\kappa \beta \alpha \lambda ,\mu \nu }\hspace{-0.3cm}&=&B_{\Sigma}^{(1)}\Delta^{\lambda \kappa }g^{[\mu \beta }g^{\nu ]\alpha }+B_{\Sigma }^{(2)} u^{\alpha }\Delta^{\lambda \kappa }u^{[\mu } \Delta^{\nu ]\beta }\nn\\&& + B_{\Sigma}^{(3)}\big(\Delta^{\lambda \kappa } \Delta^{\alpha [\mu }g^{\nu ]\beta }+\Delta^{\lambda \alpha }\Delta^{[\mu \kappa }g^{\nu ]\beta }\nn\\&&+\Delta^{\alpha \kappa }\Delta^{\lambda [\mu }g^{\nu ]\beta }\big)+B_{\Sigma}^{(4)}\Delta ^{\alpha \kappa }\Delta^{\lambda [\mu }\Delta^{\nu ]\beta}\nn\\&&+B_{\Sigma}^{(5)}u^{\alpha }\Delta^{\lambda \beta }u^{[\mu }\Delta ^{\nu ]\kappa }.
\label{eq:betaSig}
\end{eqnarray}
The explicit forms of the scalar coefficients $B^{(i)}_X$ (which are expressed by very lengthy expressions involving complicated integrals) are provided in Ref.~\cite{Bhadury:2020cop}.

It is important to emphasize that the assumption of small polarization does not introduce any new dissipative corrections to $\delta T^{\mu\nu}$ and $\delta N^\mu$ in Eqs.~\eqref{deltaN}~and~\eqref{deltaT} and their forms remain unchanged, compared to the results obtained in the usual analysis of spinless systems. Moreover, we see that the first three terms in the expression for $\delta S^{\lambda,\mu\nu}$ in Eq.~(\ref{deltaS}) arise from the same ``thermodynamic forces'' as dissipation in conserved current and energy momentum tensor. Very interestingly, the last term in Eq.~(\ref{deltaS}) leads to a new type of dissipation which is proportional to the gradient of the spin polarization tensor $\omega^{\mu\nu}$.

 We note that all the kinetic coefficients obtained from Eq.~(\ref{RTA_spin}) are proportional to the relaxation time that is common for all of them. This means that the equilibration times for momenta and spin degrees of freedom are the same. In phenomenological applications it is conceivable to vary the values of the relaxation times that appear in different kinetic coefficients, arguing that they describe independent physical phenomena. In any case, such modifications require further studies.

\section{Summary and conclusions}

In this work we have constructed kinetic theory describing evolution of the axial-current phase-space density of spin-$\onehalf$ particles in the relaxation time approximation. Our approach is based on the conservation laws for energy, linear momentum, and angular momentum. Choosing a special pseudo-gauge (by adopting the GLW version), one can split the conservation of total angular momentum into separate conservation of the orbital and spin parts. This procedure leads to the conserved spin tensor. After performing the calculations in the GLW pseudo gauge, one can switch to the canonical gauge where only the total angular momentum is conserved. We note that the pseudo-gauge transformations do not change the form of the conservation laws for the energy-momentum and angular momentum tensors. Moreover, the integrated values of energy, linear momentum and angular momentum also remain unchanged, however, they differently allocate energy density and spin density. The obtained framework has been used to derive non-equilibrium corrections to the spin tensor, which are then used to define, for the first time, the structure of spin transport coefficients in relativistic matter. 

Approaches based on the relaxation-time approximation are attractive, since they allow for a rather simple treatment of dissipative processes and very often allow for finding exact solutions. From this point of view, we expect that our formalism will be useful in studies of dissipative effects connected with spin. Clearly, more microscopic approaches are required to construct more realistic collision terms. Such problems are currently investigated by several groups, for example, in Refs.~\cite{Yang:2020hri,Li:2019qkf,Weickgenannt:2020aaf,Speranza:2020ilk,Kapusta:2020npk}.

\medskip

{\bf Acknowledgements:} S.B. and A.J. acknowledge kind hospitality of Jagiellonian University and Institute of Nuclear Physics, Krakow, where part of this work was completed. A.J. was supported in part by the DST-INSPIRE faculty award under Grant No. DST/INSPIRE/04/2017/000038. W.F. and R.R. were supported in part by the Polish National Science Centre Grants No.~2016/23/B/ST2/00717 and \\ No.~2018/30/E/ST2/00432, respectively.

\printcredits

\bibliographystyle{model6-num-names}

\bibliography{cas-refs}

\begin{thebibliography}{54}
\providecommand{\natexlab}[1]{#1}
\providecommand{\url}[1]{\texttt{#1}}
\providecommand{\href}[2]{#2}
\providecommand{\path}[1]{#1}
\providecommand{\DOIprefix}{doi:}
\providecommand{\ArXivprefix}{arXiv:}
\providecommand{\URLprefix}{URL: }
\providecommand{\Pubmedprefix}{pmid:}
\providecommand{\doi}[1]{\href{http://dx.doi.org/#1}{\path{#1}}}
\providecommand{\Pubmed}[1]{\href{pmid:#1}{\path{#1}}}
\providecommand{\BIBand}{and}
\providecommand{\bibinfo}[2]{#2}
\ifx\xfnm\undefined \def\xfnm[#1]{\unskip,\space#1}\fi
\makeatletter\def\@biblabel#1{#1.}\makeatother
\bibitem[{Adamczyk et~al.(2017)}]{STAR:2017ckg}
\bibinfo{author}{Adamczyk\xfnm[ L.]}, et~al. (\bibinfo{collaboration}{STAR}).
\newblock \bibinfo{title}{{Global $\Lambda$ hyperon polarization in nuclear
  collisions: evidence for the most vortical fluid}}.
\newblock \emph{\bibinfo{journal}{Nature}}
  \bibinfo{year}{2017};\bibinfo{volume}{548}:\bibinfo{pages}{62--65}.
\newblock \DOIprefix\doi{10.1038/nature23004}.
  \href{http://arxiv.org/abs/1701.06657}{\tt arXiv:1701.06657}.
\bibitem[{Adam et~al.(2018)}]{Adam:2018ivw}
\bibinfo{author}{Adam\xfnm[ J.]}, et~al. (\bibinfo{collaboration}{STAR}).
\newblock \bibinfo{title}{{Global polarization of $\Lambda$ hyperons in Au+Au
  collisions at $\sqrt{s_{_{NN}}}$ = 200 GeV}}.
\newblock \emph{\bibinfo{journal}{Phys Rev}}
  \bibinfo{year}{2018};\bibinfo{volume}{C98}:\bibinfo{pages}{014910}.
\newblock \DOIprefix\doi{10.1103/PhysRevC.98.014910}.
  \href{http://arxiv.org/abs/1805.04400}{\tt arXiv:1805.04400}.
\bibitem[{Acharya et~al.(2020)}]{Acharya:2019vpe}
\bibinfo{author}{Acharya\xfnm[ S.]}, et~al. (\bibinfo{collaboration}{ALICE}).
\newblock \bibinfo{title}{{Measurement of spin-orbital angular momentum
  interactions in relativistic heavy-ion collisions}}.
\newblock \emph{\bibinfo{journal}{Phys Rev Lett}}
  \bibinfo{year}{2020};\bibinfo{volume}{125}(\bibinfo{number}{1}):\bibinfo{pages}{012301}.
\newblock \DOIprefix\doi{10.1103/PhysRevLett.125.012301}.
  \href{http://arxiv.org/abs/1910.14408}{\tt arXiv:1910.14408}.
\bibitem[{Liang and Wang(2005)}]{Liang:2004ph}
\bibinfo{author}{Liang\xfnm[ Z.T.]}, \bibinfo{author}{Wang\xfnm[ X.N.]}.
\newblock \bibinfo{title}{{Globally polarized quark-gluon plasma in non-central
  A+A collisions}}.
\newblock \emph{\bibinfo{journal}{Phys Rev Lett}}
  \bibinfo{year}{2005};\bibinfo{volume}{94}:\bibinfo{pages}{102301}.
\newblock \DOIprefix\doi{10.1103/PhysRevLett.94.102301,
  10.1103/PhysRevLett.96.039901}.
  \href{http://arxiv.org/abs/nucl-th/0410079}{\tt arXiv:nucl-th/0410079};
  \bibinfo{note}{[Erratum: Phys. Rev. Lett.96,039901(2006)]}.
\bibitem[{Betz et~al.(2007)Betz, Gyulassy and Torrieri}]{Betz:2007kg}
\bibinfo{author}{Betz\xfnm[ B.]}, \bibinfo{author}{Gyulassy\xfnm[ M.]},
  \bibinfo{author}{Torrieri\xfnm[ G.]}.
\newblock \bibinfo{title}{{Polarization probes of vorticity in heavy ion
  collisions}}.
\newblock \emph{\bibinfo{journal}{Phys Rev}}
  \bibinfo{year}{2007};\bibinfo{volume}{C76}:\bibinfo{pages}{044901}.
\newblock \DOIprefix\doi{10.1103/PhysRevC.76.044901}.
  \href{http://arxiv.org/abs/0708.0035}{\tt arXiv:0708.0035}.
\bibitem[{Voloshin(2004)}]{Voloshin:2004ha}
\bibinfo{author}{Voloshin\xfnm[ S.A.]}.
\newblock \bibinfo{title}{{Polarized secondary particles in unpolarized high
  energy hadron-hadron collisions?}}
  \bibinfo{year}{2004};\href{http://arxiv.org/abs/nucl-th/0410089}{\tt
  arXiv:nucl-th/0410089}.
\bibitem[{Becattini and Piccinini(2008)}]{Becattini:2007nd}
\bibinfo{author}{Becattini\xfnm[ F.]}, \bibinfo{author}{Piccinini\xfnm[ F.]}.
\newblock \bibinfo{title}{{The Ideal relativistic spinning gas: Polarization
  and spectra}}.
\newblock \emph{\bibinfo{journal}{Annals Phys}}
  \bibinfo{year}{2008};\bibinfo{volume}{323}:\bibinfo{pages}{2452--2473}.
\newblock \DOIprefix\doi{10.1016/j.aop.2008.01.001}.
  \href{http://arxiv.org/abs/0710.5694}{\tt arXiv:0710.5694}.
\bibitem[{Becattini et~al.(2008)Becattini, Piccinini and
  Rizzo}]{Becattini:2007sr}
\bibinfo{author}{Becattini\xfnm[ F.]}, \bibinfo{author}{Piccinini\xfnm[ F.]},
  \bibinfo{author}{Rizzo\xfnm[ J.]}.
\newblock \bibinfo{title}{{Angular momentum conservation in heavy ion
  collisions at very high energy}}.
\newblock \emph{\bibinfo{journal}{Phys Rev}}
  \bibinfo{year}{2008};\bibinfo{volume}{C77}:\bibinfo{pages}{024906}.
\newblock \DOIprefix\doi{10.1103/PhysRevC.77.024906}.
  \href{http://arxiv.org/abs/0711.1253}{\tt arXiv:0711.1253}.
\bibitem[{Becattini et~al.(2013{\natexlab{a}})Becattini, Chandra, Del~Zanna and
  Grossi}]{Becattini:2013fla}
\bibinfo{author}{Becattini\xfnm[ F.]}, \bibinfo{author}{Chandra\xfnm[ V.]},
  \bibinfo{author}{Del~Zanna\xfnm[ L.]}, \bibinfo{author}{Grossi\xfnm[ E.]}.
\newblock \bibinfo{title}{{Relativistic distribution function for particles
  with spin at local thermodynamical equilibrium}}.
\newblock \emph{\bibinfo{journal}{Annals Phys}}
  \bibinfo{year}{2013}{\natexlab{a}};\bibinfo{volume}{338}:\bibinfo{pages}{32--49}.
\newblock \DOIprefix\doi{10.1016/j.aop.2013.07.004}.
  \href{http://arxiv.org/abs/1303.3431}{\tt arXiv:1303.3431}.
\bibitem[{Becattini et~al.(2013{\natexlab{b}})Becattini, Csernai and
  Wang}]{Becattini:2013vja}
\bibinfo{author}{Becattini\xfnm[ F.]}, \bibinfo{author}{Csernai\xfnm[ L.]},
  \bibinfo{author}{Wang\xfnm[ D.J.]}.
\newblock \bibinfo{title}{{$\Lambda$ polarization in peripheral heavy ion
  collisions}}.
\newblock \emph{\bibinfo{journal}{Phys Rev}}
  \bibinfo{year}{2013}{\natexlab{b}};\bibinfo{volume}{C88}(\bibinfo{number}{3}):\bibinfo{pages}{034905}.
\newblock \DOIprefix\doi{10.1103/PhysRevC.93.069901,
  10.1103/PhysRevC.88.034905}. \href{http://arxiv.org/abs/1304.4427}{\tt
  arXiv:1304.4427}; \bibinfo{note}{[Erratum: Phys. Rev.C93,no.6,069901(2016)]}.
\bibitem[{Li et~al.(2017)Li, Pang, Wang and Xia}]{Li:2017slc}
\bibinfo{author}{Li\xfnm[ H.]}, \bibinfo{author}{Pang\xfnm[ L.G.]},
  \bibinfo{author}{Wang\xfnm[ Q.]}, \bibinfo{author}{Xia\xfnm[ X.L.]}.
\newblock \bibinfo{title}{{Global $\Lambda$ polarization in heavy-ion
  collisions from a transport model}}.
\newblock \emph{\bibinfo{journal}{Phys Rev}}
  \bibinfo{year}{2017};\bibinfo{volume}{C96}(\bibinfo{number}{5}):\bibinfo{pages}{054908}.
\newblock \DOIprefix\doi{10.1103/PhysRevC.96.054908}.
  \href{http://arxiv.org/abs/1704.01507}{\tt arXiv:1704.01507}.
\bibitem[{Sun and Ko(2017)}]{Sun:2017xhx}
\bibinfo{author}{Sun\xfnm[ Y.]}, \bibinfo{author}{Ko\xfnm[ C.M.]}.
\newblock \bibinfo{title}{{$\Lambda$ hyperon polarization in relativistic heavy
  ion collisions from a chiral kinetic approach}}.
\newblock \emph{\bibinfo{journal}{Phys Rev C}}
  \bibinfo{year}{2017};\bibinfo{volume}{96}(\bibinfo{number}{2}):\bibinfo{pages}{024906}.
\newblock \DOIprefix\doi{10.1103/PhysRevC.96.024906}.
  \href{http://arxiv.org/abs/1706.09467}{\tt arXiv:1706.09467}.
\bibitem[{Niida(2019)}]{Niida:2018hfw}
\bibinfo{author}{Niida\xfnm[ T.]} (\bibinfo{collaboration}{STAR}).
\newblock \bibinfo{title}{{Global and local polarization of $\Lambda$ hyperons
  in Au+Au collisions at 200 GeV from STAR}}.
\newblock \emph{\bibinfo{journal}{Nucl Phys}}
  \bibinfo{year}{2019};\bibinfo{volume}{A982}:\bibinfo{pages}{511--514}.
\newblock \DOIprefix\doi{10.1016/j.nuclphysa.2018.08.034}.
  \href{http://arxiv.org/abs/1808.10482}{\tt arXiv:1808.10482}.
\bibitem[{Adam et~al.(2019)}]{Adam:2019srw}
\bibinfo{author}{Adam\xfnm[ J.]}, et~al. (\bibinfo{collaboration}{STAR}).
\newblock \bibinfo{title}{{Polarization of $\Lambda$ ($\bar{\Lambda}$) hyperons
  along the beam direction in Au+Au collisions at $\sqrt{s_{_{NN}}}$ = 200
  GeV}}.
\newblock \emph{\bibinfo{journal}{Phys Rev Lett}}
  \bibinfo{year}{2019};\bibinfo{volume}{123}(\bibinfo{number}{13}):\bibinfo{pages}{132301}.
\newblock \DOIprefix\doi{10.1103/PhysRevLett.123.132301}.
  \href{http://arxiv.org/abs/1905.11917}{\tt arXiv:1905.11917}.
\bibitem[{Lisa(hina)}]{Lisa:2019}
\bibinfo{author}{Lisa\xfnm[ M.]}.
\newblock \bibinfo{title}{{Polarization/Alignment $\&$ Charge-Separation:
  Experimental Status}}.
\newblock \bibinfo{year}{talk given at Quark Matter 2019, Wuhan, China}.
\bibitem[{Becattini et~al.(2019{\natexlab{a}})Becattini, Florkowski and
  Speranza}]{Becattini:2018duy}
\bibinfo{author}{Becattini\xfnm[ F.]}, \bibinfo{author}{Florkowski\xfnm[ W.]},
  \bibinfo{author}{Speranza\xfnm[ E.]}.
\newblock \bibinfo{title}{{Spin tensor and its role in non-equilibrium
  thermodynamics}}.
\newblock \emph{\bibinfo{journal}{Phys Lett}}
  \bibinfo{year}{2019}{\natexlab{a}};\bibinfo{volume}{B789}:\bibinfo{pages}{419--425}.
\newblock \DOIprefix\doi{10.1016/j.physletb.2018.12.016}.
  \href{http://arxiv.org/abs/1807.10994}{\tt arXiv:1807.10994}.
\bibitem[{Hattori et~al.(2019)Hattori, Hongo, Huang, Matsuo and
  Taya}]{Hattori:2019lfp}
\bibinfo{author}{Hattori\xfnm[ K.]}, \bibinfo{author}{Hongo\xfnm[ M.]},
  \bibinfo{author}{Huang\xfnm[ X.G.]}, \bibinfo{author}{Matsuo\xfnm[ M.]},
  \bibinfo{author}{Taya\xfnm[ H.]}.
\newblock \bibinfo{title}{{Fate of spin polarization in a relativistic fluid:
  An entropy-current analysis}}.
\newblock \emph{\bibinfo{journal}{Phys Lett}}
  \bibinfo{year}{2019};\bibinfo{volume}{B795}:\bibinfo{pages}{100--106}.
\newblock \DOIprefix\doi{10.1016/j.physletb.2019.05.040}.
  \href{http://arxiv.org/abs/1901.06615}{\tt arXiv:1901.06615}.
\bibitem[{Wu et~al.(2019)Wu, Pang, Huang and Wang}]{Wu:2019eyi}
\bibinfo{author}{Wu\xfnm[ H.Z.]}, \bibinfo{author}{Pang\xfnm[ L.G.]},
  \bibinfo{author}{Huang\xfnm[ X.G.]}, \bibinfo{author}{Wang\xfnm[ Q.]}.
\newblock \bibinfo{title}{{Local spin polarization in high energy heavy ion
  collisions}}.
\newblock \emph{\bibinfo{journal}{Phys Rev Research}}
  \bibinfo{year}{2019};\bibinfo{volume}{1}:\bibinfo{pages}{033058}.
\newblock \DOIprefix\doi{10.1103/PhysRevResearch.1.033058}.
  \href{http://arxiv.org/abs/1906.09385}{\tt arXiv:1906.09385}.
\bibitem[{Sheng et~al.(2020)Sheng, Oliva and Wang}]{Sheng:2019kmk}
\bibinfo{author}{Sheng\xfnm[ X.L.]}, \bibinfo{author}{Oliva\xfnm[ L.]},
  \bibinfo{author}{Wang\xfnm[ Q.]}.
\newblock \bibinfo{title}{{What can we learn from the global spin alignment of
  $\phi$ mesons in heavy-ion collisions?}}
\newblock \emph{\bibinfo{journal}{Phys Rev D}}
  \bibinfo{year}{2020};\bibinfo{volume}{101}(\bibinfo{number}{9}):\bibinfo{pages}{096005}.
\newblock \DOIprefix\doi{10.1103/PhysRevD.101.096005}.
  \href{http://arxiv.org/abs/1910.13684}{\tt arXiv:1910.13684}.
\bibitem[{Becattini et~al.(2019{\natexlab{b}})Becattini, Cao and
  Speranza}]{Becattini:2019ntv}
\bibinfo{author}{Becattini\xfnm[ F.]}, \bibinfo{author}{Cao\xfnm[ G.]},
  \bibinfo{author}{Speranza\xfnm[ E.]}.
\newblock \bibinfo{title}{{Polarization transfer in hyperon decays and its
  effect in relativistic nuclear collisions}}.
\newblock \emph{\bibinfo{journal}{Eur Phys J C}}
  \bibinfo{year}{2019}{\natexlab{b}};\bibinfo{volume}{79}(\bibinfo{number}{9}):\bibinfo{pages}{741}.
\newblock \DOIprefix\doi{10.1140/epjc/s10052-019-7213-6}.
  \href{http://arxiv.org/abs/1905.03123}{\tt arXiv:1905.03123}.
\bibitem[{Liu et~al.(2020)Liu, Sun and Ko}]{Liu:2019krs}
\bibinfo{author}{Liu\xfnm[ S.Y.]}, \bibinfo{author}{Sun\xfnm[ Y.]},
  \bibinfo{author}{Ko\xfnm[ C.M.]}.
\newblock \bibinfo{title}{{Spin Polarizations in a Covariant
  Angular-Momentum-Conserved Chiral Transport Model}}.
\newblock \emph{\bibinfo{journal}{Phys Rev Lett}}
  \bibinfo{year}{2020};\bibinfo{volume}{125}(\bibinfo{number}{6}):\bibinfo{pages}{062301}.
\newblock \DOIprefix\doi{10.1103/PhysRevLett.125.062301}.
  \href{http://arxiv.org/abs/1910.06774}{\tt arXiv:1910.06774}.
\bibitem[{Yang et~al.(2020)Yang, Hattori and Hidaka}]{Yang:2020hri}
\bibinfo{author}{Yang\xfnm[ D.L.]}, \bibinfo{author}{Hattori\xfnm[ K.]},
  \bibinfo{author}{Hidaka\xfnm[ Y.]}.
\newblock \bibinfo{title}{{Effective quantum kinetic theory for spin transport
  of fermions with collsional effects}}.
\newblock \emph{\bibinfo{journal}{JHEP}}
  \bibinfo{year}{2020};\bibinfo{volume}{20}:\bibinfo{pages}{070}.
\newblock \DOIprefix\doi{10.1007/JHEP07(2020)070}.
  \href{http://arxiv.org/abs/2002.02612}{\tt arXiv:2002.02612}.
\bibitem[{Gallegos and Gürsoy(2020)}]{Gallegos:2020otk}
\bibinfo{author}{Gallegos\xfnm[ A.]}, \bibinfo{author}{Gürsoy\xfnm[ U.]}.
\newblock \bibinfo{title}{{Holographic spin liquids and Lovelock Chern-Simons
  gravity}} \bibinfo{year}{2020};\href{http://arxiv.org/abs/2004.05148}{\tt
  arXiv:2004.05148}.
\bibitem[{Florkowski et~al.(2018{\natexlab{a}})Florkowski, Friman, Jaiswal and
  Speranza}]{Florkowski:2017ruc}
\bibinfo{author}{Florkowski\xfnm[ W.]}, \bibinfo{author}{Friman\xfnm[ B.]},
  \bibinfo{author}{Jaiswal\xfnm[ A.]}, \bibinfo{author}{Speranza\xfnm[ E.]}.
\newblock \bibinfo{title}{{Relativistic fluid dynamics with spin}}.
\newblock \emph{\bibinfo{journal}{Phys Rev}}
  \bibinfo{year}{2018}{\natexlab{a}};\bibinfo{volume}{C97}(\bibinfo{number}{4}):\bibinfo{pages}{041901}.
\newblock \DOIprefix\doi{10.1103/PhysRevC.97.041901}.
  \href{http://arxiv.org/abs/1705.00587}{\tt arXiv:1705.00587}.
\bibitem[{Florkowski et~al.(2018{\natexlab{b}})Florkowski, Kumar and
  Ryblewski}]{Florkowski:2018ahw}
\bibinfo{author}{Florkowski\xfnm[ W.]}, \bibinfo{author}{Kumar\xfnm[ A.]},
  \bibinfo{author}{Ryblewski\xfnm[ R.]}.
\newblock \bibinfo{title}{{Thermodynamic versus kinetic approach to
  polarization-vorticity coupling}}.
\newblock \emph{\bibinfo{journal}{Phys Rev}}
  \bibinfo{year}{2018}{\natexlab{b}};\bibinfo{volume}{C98}:\bibinfo{pages}{044906}.
\newblock \DOIprefix\doi{10.1103/PhysRevC.98.044906}.
  \href{http://arxiv.org/abs/1806.02616}{\tt arXiv:1806.02616}.
\bibitem[{Florkowski et~al.(2019)Florkowski, Kumar and
  Ryblewski}]{Florkowski:2018fap}
\bibinfo{author}{Florkowski\xfnm[ W.]}, \bibinfo{author}{Kumar\xfnm[ A.]},
  \bibinfo{author}{Ryblewski\xfnm[ R.]}.
\newblock \bibinfo{title}{{Relativistic hydrodynamics for spin-polarized
  fluids}}.
\newblock \emph{\bibinfo{journal}{Prog Part Nucl Phys}}
  \bibinfo{year}{2019};\bibinfo{volume}{108}:\bibinfo{pages}{103709}.
\newblock \DOIprefix\doi{10.1016/j.ppnp.2019.07.001}.
  \href{http://arxiv.org/abs/1811.04409}{\tt arXiv:1811.04409}.
\bibitem[{Bhatnagar et~al.(1954)Bhatnagar, Gross and Krook}]{PhysRev.94.511}
\bibinfo{author}{Bhatnagar\xfnm[ P.L.]}, \bibinfo{author}{Gross\xfnm[ E.P.]},
  \bibinfo{author}{Krook\xfnm[ M.]}.
\newblock \bibinfo{title}{A model for collision processes in gases. i. small
  amplitude processes in charged and neutral one-component systems}.
\newblock \emph{\bibinfo{journal}{Phys Rev}}
  \bibinfo{year}{1954};\bibinfo{volume}{94}:\bibinfo{pages}{511--525}.
\newblock \URLprefix \url{https://link.aps.org/doi/10.1103/PhysRev.94.511}.
  \DOIprefix\doi{10.1103/PhysRev.94.511}.
\bibitem[{Baym(1984)}]{Baym:1984np}
\bibinfo{author}{Baym\xfnm[ G.]}.
\newblock \bibinfo{title}{{THERMAL EQUILIBRATION IN ULTRARELATIVISTIC HEAVY ION
  COLLISIONS}}.
\newblock \emph{\bibinfo{journal}{Phys Lett B}}
  \bibinfo{year}{1984};\bibinfo{volume}{138}:\bibinfo{pages}{18--22}.
\newblock \DOIprefix\doi{10.1016/0370-2693(84)91863-X}.
\bibitem[{Florkowski et~al.(2013)Florkowski, Ryblewski and
  Strickland}]{Florkowski:2013lya}
\bibinfo{author}{Florkowski\xfnm[ W.]}, \bibinfo{author}{Ryblewski\xfnm[ R.]},
  \bibinfo{author}{Strickland\xfnm[ M.]}.
\newblock \bibinfo{title}{{Testing viscous and anisotropic hydrodynamics in an
  exactly solvable case}}.
\newblock \emph{\bibinfo{journal}{Phys Rev C}}
  \bibinfo{year}{2013};\bibinfo{volume}{88}:\bibinfo{pages}{024903}.
\newblock \DOIprefix\doi{10.1103/PhysRevC.88.024903}.
  \href{http://arxiv.org/abs/1305.7234}{\tt arXiv:1305.7234}.
\bibitem[{Denicol et~al.(2014)Denicol, Heinz, Martinez, Noronha and
  Strickland}]{Denicol:2014xca}
\bibinfo{author}{Denicol\xfnm[ G.S.]}, \bibinfo{author}{Heinz\xfnm[ U.W.]},
  \bibinfo{author}{Martinez\xfnm[ M.]}, \bibinfo{author}{Noronha\xfnm[ J.]},
  \bibinfo{author}{Strickland\xfnm[ M.]}.
\newblock \bibinfo{title}{{New Exact Solution of the Relativistic Boltzmann
  Equation and its Hydrodynamic Limit}}.
\newblock \emph{\bibinfo{journal}{Phys Rev Lett}}
  \bibinfo{year}{2014};\bibinfo{volume}{113}(\bibinfo{number}{20}):\bibinfo{pages}{202301}.
\newblock \DOIprefix\doi{10.1103/PhysRevLett.113.202301}.
  \href{http://arxiv.org/abs/1408.5646}{\tt arXiv:1408.5646}.
\bibitem[{Heller and Spalinski(2015)}]{Heller:2015dha}
\bibinfo{author}{Heller\xfnm[ M.P.]}, \bibinfo{author}{Spalinski\xfnm[ M.]}.
\newblock \bibinfo{title}{{Hydrodynamics Beyond the Gradient Expansion:
  Resurgence and Resummation}}.
\newblock \emph{\bibinfo{journal}{Phys Rev Lett}}
  \bibinfo{year}{2015};\bibinfo{volume}{115}(\bibinfo{number}{7}):\bibinfo{pages}{072501}.
\newblock \DOIprefix\doi{10.1103/PhysRevLett.115.072501}.
  \href{http://arxiv.org/abs/1503.07514}{\tt arXiv:1503.07514}.
\bibitem[{Heller et~al.(2018)Heller, Kurkela, Spali\'nski and
  Svensson}]{Heller:2016rtz}
\bibinfo{author}{Heller\xfnm[ M.P.]}, \bibinfo{author}{Kurkela\xfnm[ A.]},
  \bibinfo{author}{Spali\'nski\xfnm[ M.]}, \bibinfo{author}{Svensson\xfnm[
  V.]}.
\newblock \bibinfo{title}{{Hydrodynamization in kinetic theory: Transient modes
  and the gradient expansion}}.
\newblock \emph{\bibinfo{journal}{Phys Rev D}}
  \bibinfo{year}{2018};\bibinfo{volume}{97}(\bibinfo{number}{9}):\bibinfo{pages}{091503}.
\newblock \DOIprefix\doi{10.1103/PhysRevD.97.091503}.
  \href{http://arxiv.org/abs/1609.04803}{\tt arXiv:1609.04803}.
\bibitem[{Romatschke(2018)}]{Romatschke:2017vte}
\bibinfo{author}{Romatschke\xfnm[ P.]}.
\newblock \bibinfo{title}{{Relativistic Fluid Dynamics Far From Local
  Equilibrium}}.
\newblock \emph{\bibinfo{journal}{Phys Rev Lett}}
  \bibinfo{year}{2018};\bibinfo{volume}{120}(\bibinfo{number}{1}):\bibinfo{pages}{012301}.
\newblock \DOIprefix\doi{10.1103/PhysRevLett.120.012301}.
  \href{http://arxiv.org/abs/1704.08699}{\tt arXiv:1704.08699}.
\bibitem[{Florkowski et~al.(2018{\natexlab{c}})Florkowski, Friman, Jaiswal,
  Ryblewski and Speranza}]{Florkowski:2017dyn}
\bibinfo{author}{Florkowski\xfnm[ W.]}, \bibinfo{author}{Friman\xfnm[ B.]},
  \bibinfo{author}{Jaiswal\xfnm[ A.]}, \bibinfo{author}{Ryblewski\xfnm[ R.]},
  \bibinfo{author}{Speranza\xfnm[ E.]}.
\newblock \bibinfo{title}{{Spin-dependent distribution functions for
  relativistic hydrodynamics of spin-1/2 particles}}.
\newblock \emph{\bibinfo{journal}{Phys Rev}}
  \bibinfo{year}{2018}{\natexlab{c}};\bibinfo{volume}{D97}(\bibinfo{number}{11}):\bibinfo{pages}{116017}.
\newblock \DOIprefix\doi{10.1103/PhysRevD.97.116017}.
  \href{http://arxiv.org/abs/1712.07676}{\tt arXiv:1712.07676}.
\bibitem[{Leader(2011)}]{Leader:2001gr}
\bibinfo{author}{Leader\xfnm[ E.]}.
\newblock \bibinfo{title}{{Spin in particle physics}}.
\newblock \emph{\bibinfo{journal}{Camb Monogr Part Phys Nucl Phys Cosmol}}
  \bibinfo{year}{2011};\bibinfo{volume}{15}:\bibinfo{pages}{pp.1--500}.
\bibitem[{Vasak et~al.(1987)Vasak, Gyulassy and Elze}]{Vasak:1987um}
\bibinfo{author}{Vasak\xfnm[ D.]}, \bibinfo{author}{Gyulassy\xfnm[ M.]},
  \bibinfo{author}{Elze\xfnm[ H.T.]}.
\newblock \bibinfo{title}{{Quantum Transport Theory for Abelian Plasmas}}.
\newblock \emph{\bibinfo{journal}{Annals Phys}}
  \bibinfo{year}{1987};\bibinfo{volume}{173}:\bibinfo{pages}{462--492}.
\newblock \DOIprefix\doi{10.1016/0003-4916(87)90169-2}.
\bibitem[{Zhuang and Heinz(1996)}]{Zhuang:1995pd}
\bibinfo{author}{Zhuang\xfnm[ P.]}, \bibinfo{author}{Heinz\xfnm[ U.W.]}.
\newblock \bibinfo{title}{{Relativistic quantum transport theory for
  electrodynamics}}.
\newblock \emph{\bibinfo{journal}{Annals Phys}}
  \bibinfo{year}{1996};\bibinfo{volume}{245}:\bibinfo{pages}{311--338}.
\newblock \DOIprefix\doi{10.1006/aphy.1996.0011}.
  \href{http://arxiv.org/abs/nucl-th/9502034}{\tt arXiv:nucl-th/9502034}.
\bibitem[{Elze et~al.(1986{\natexlab{a}})Elze, Gyulassy and
  Vasak}]{Elze:1986hq}
\bibinfo{author}{Elze\xfnm[ H.T.]}, \bibinfo{author}{Gyulassy\xfnm[ M.]},
  \bibinfo{author}{Vasak\xfnm[ D.]}.
\newblock \bibinfo{title}{{Transport Equations for the {QCD} Gluon Wigner
  Operator}}.
\newblock \emph{\bibinfo{journal}{Phys Lett}}
  \bibinfo{year}{1986}{\natexlab{a}};\bibinfo{volume}{B177}:\bibinfo{pages}{402--408}.
\newblock \DOIprefix\doi{10.1016/0370-2693(86)90778-1}.
\bibitem[{Elze et~al.(1986{\natexlab{b}})Elze, Gyulassy and
  Vasak}]{Elze:1986qd}
\bibinfo{author}{Elze\xfnm[ H.T.]}, \bibinfo{author}{Gyulassy\xfnm[ M.]},
  \bibinfo{author}{Vasak\xfnm[ D.]}.
\newblock \bibinfo{title}{{Transport Equations for the {QCD} Quark Wigner
  Operator}}.
\newblock \emph{\bibinfo{journal}{Nucl Phys}}
  \bibinfo{year}{1986}{\natexlab{b}};\bibinfo{volume}{B276}:\bibinfo{pages}{706--728}.
\newblock \DOIprefix\doi{10.1016/0550-3213(86)90072-6}.
\bibitem[{Florkowski et~al.(1996)Florkowski, Hufner, Klevansky and
  Neise}]{Florkowski:1995ei}
\bibinfo{author}{Florkowski\xfnm[ W.]}, \bibinfo{author}{Hufner\xfnm[ J.]},
  \bibinfo{author}{Klevansky\xfnm[ S.P.]}, \bibinfo{author}{Neise\xfnm[ L.]}.
\newblock \bibinfo{title}{{Chirally invariant transport equations for quark
  matter}}.
\newblock \emph{\bibinfo{journal}{Annals Phys}}
  \bibinfo{year}{1996};\bibinfo{volume}{245}:\bibinfo{pages}{445--463}.
\newblock \DOIprefix\doi{10.1006/aphy.1996.0016}.
  \href{http://arxiv.org/abs/hep-ph/9505407}{\tt arXiv:hep-ph/9505407}.
\bibitem[{Weickgenannt et~al.(2019)Weickgenannt, Sheng, Speranza, Wang and
  Rischke}]{Weickgenannt:2019dks}
\bibinfo{author}{Weickgenannt\xfnm[ N.]}, \bibinfo{author}{Sheng\xfnm[ X.L.]},
  \bibinfo{author}{Speranza\xfnm[ E.]}, \bibinfo{author}{Wang\xfnm[ Q.]},
  \bibinfo{author}{Rischke\xfnm[ D.H.]}.
\newblock \bibinfo{title}{{Kinetic theory for massive spin-1/2 particles from
  the Wigner-function formalism}}.
\newblock \emph{\bibinfo{journal}{Phys Rev D}}
  \bibinfo{year}{2019};\bibinfo{volume}{100}(\bibinfo{number}{5}):\bibinfo{pages}{056018}.
\newblock \DOIprefix\doi{10.1103/PhysRevD.100.056018}.
  \href{http://arxiv.org/abs/1902.06513}{\tt arXiv:1902.06513}.
\bibitem[{Florkowski et~al.(2020)Florkowski, Kumar and
  Ryblewski}]{Florkowski:2019gio}
\bibinfo{author}{Florkowski\xfnm[ W.]}, \bibinfo{author}{Kumar\xfnm[ A.]},
  \bibinfo{author}{Ryblewski\xfnm[ R.]}.
\newblock \bibinfo{title}{{Spin Potential for Relativistic Particles with Spin
  1/2}}.
\newblock \emph{\bibinfo{journal}{Acta Phys Polon B}}
  \bibinfo{year}{2020};\bibinfo{volume}{51}:\bibinfo{pages}{945--959}.
\newblock \DOIprefix\doi{10.5506/APhysPolB.51.945}.
  \href{http://arxiv.org/abs/1907.09835}{\tt arXiv:1907.09835}.
\bibitem[{Prokhorov and Teryaev(2018)}]{Prokhorov:2017atp}
\bibinfo{author}{Prokhorov\xfnm[ G.]}, \bibinfo{author}{Teryaev\xfnm[ O.]}.
\newblock \bibinfo{title}{{Anomalous current from the covariant Wigner
  function}}.
\newblock \emph{\bibinfo{journal}{Phys Rev D}}
  \bibinfo{year}{2018};\bibinfo{volume}{97}(\bibinfo{number}{7}):\bibinfo{pages}{076013}.
\newblock \DOIprefix\doi{10.1103/PhysRevD.97.076013}.
  \href{http://arxiv.org/abs/1707.02491}{\tt arXiv:1707.02491}.
\bibitem[{{Anderson} and {Witting}(1974)}]{anderson1974relativistic}
\bibinfo{author}{{Anderson}\xfnm[ J.L.]}, \bibinfo{author}{{Witting}\xfnm[
  H.R.]}.
\newblock \bibinfo{title}{{A relativistic relaxation-time model for the
  Boltzmann equation}}.
\newblock \emph{\bibinfo{journal}{Physica}}
  \bibinfo{year}{1974};\bibinfo{volume}{74}(\bibinfo{number}{3}):\bibinfo{pages}{466--488}.
\newblock \DOIprefix\doi{10.1016/0031-8914(74)90355-3}.
\bibitem[{Fradkin and Good(1961)}]{Fradkin:1961}
\bibinfo{author}{Fradkin\xfnm[ D.M.]}, \bibinfo{author}{Good\xfnm[ R.H.]}.
\newblock \bibinfo{title}{{Tensor Operator for Electron Polarization}}.
\newblock \emph{\bibinfo{journal}{Il Nuovo Cimento}}
  \bibinfo{year}{1961};\bibinfo{volume}{22}:\bibinfo{pages}{643}.
\bibitem[{Hilgevoord and Wouthuysen(1962)}]{Hilgevoord:1962}
\bibinfo{author}{Hilgevoord\xfnm[ J.]}, \bibinfo{author}{Wouthuysen\xfnm[
  S.A.]}.
\newblock \bibinfo{title}{{On the spin angular momentum of the Dirac
  particle}}.
\newblock \emph{\bibinfo{journal}{Nucl Phys A}}
  \bibinfo{year}{1962};\bibinfo{volume}{40}:\bibinfo{pages}{1}.
\bibitem[{De~Groot(1980)}]{DeGroot:1980dk}
\bibinfo{author}{De~Groot\xfnm[ S.R.]}.
\newblock \bibinfo{title}{{Relativistic Kinetic Theory. Principles and
  Applications}}.
\newblock \bibinfo{year}{1980}.
\bibitem[{Hess and Waldmann(1966)}]{Hess:1966}
\bibinfo{author}{Hess\xfnm[ S.]}, \bibinfo{author}{Waldmann\xfnm[ L.]}.
\newblock \emph{\bibinfo{journal}{Zeitschrift fuer Naturforschung}}
  \bibinfo{year}{1966};\bibinfo{volume}{21a}:\bibinfo{pages}{1529}.
\bibitem[{Mathisson(1937)}]{Mathisson:1937zz}
\bibinfo{author}{Mathisson\xfnm[ M.]}.
\newblock \bibinfo{title}{{Neue mechanik materieller systemes}}.
\newblock \emph{\bibinfo{journal}{Acta Phys Polon}}
  \bibinfo{year}{1937};\bibinfo{volume}{6}:\bibinfo{pages}{163--2900}.
\bibitem[{Bhadury et~al.(2020)Bhadury, Florkowski, Jaiswal, Kumar and
  Ryblewski}]{Bhadury:2020cop}
\bibinfo{author}{Bhadury\xfnm[ S.]}, \bibinfo{author}{Florkowski\xfnm[ W.]},
  \bibinfo{author}{Jaiswal\xfnm[ A.]}, \bibinfo{author}{Kumar\xfnm[ A.]},
  \bibinfo{author}{Ryblewski\xfnm[ R.]}.
\newblock \bibinfo{title}{{Dissipative Spin Dynamics in Relativistic Matter}}
  \bibinfo{year}{2020};\href{http://arxiv.org/abs/2008.10976}{\tt
  arXiv:2008.10976}.
\bibitem[{Li and Yee(2019)}]{Li:2019qkf}
\bibinfo{author}{Li\xfnm[ S.]}, \bibinfo{author}{Yee\xfnm[ H.U.]}.
\newblock \bibinfo{title}{{Quantum Kinetic Theory of Spin Polarization of
  Massive Quarks in Perturbative QCD: Leading Log}}.
\newblock \emph{\bibinfo{journal}{Phys Rev D}}
  \bibinfo{year}{2019};\bibinfo{volume}{100}(\bibinfo{number}{5}):\bibinfo{pages}{056022}.
\newblock \DOIprefix\doi{10.1103/PhysRevD.100.056022}.
  \href{http://arxiv.org/abs/1905.10463}{\tt arXiv:1905.10463}.
\bibitem[{Weickgenannt et~al.(2020)Weickgenannt, Speranza, Sheng, Wang and
  Rischke}]{Weickgenannt:2020aaf}
\bibinfo{author}{Weickgenannt\xfnm[ N.]}, \bibinfo{author}{Speranza\xfnm[ E.]},
  \bibinfo{author}{Sheng\xfnm[ X.l.]}, \bibinfo{author}{Wang\xfnm[ Q.]},
  \bibinfo{author}{Rischke\xfnm[ D.H.]}.
\newblock \bibinfo{title}{{Generating spin polarization from vorticity through
  nonlocal collisions}}
  \bibinfo{year}{2020};\href{http://arxiv.org/abs/2005.01506}{\tt
  arXiv:2005.01506}.
\bibitem[{Speranza and Weickgenannt(2020)}]{Speranza:2020ilk}
\bibinfo{author}{Speranza\xfnm[ E.]}, \bibinfo{author}{Weickgenannt\xfnm[ N.]}.
\newblock \bibinfo{title}{{Spin tensor and pseudo-gauges: from nuclear
  collisions to gravitational physics}}
  \bibinfo{year}{2020};\href{http://arxiv.org/abs/2007.00138}{\tt
  arXiv:2007.00138}.
\bibitem[{Kapusta et~al.(2020)Kapusta, Rrapaj and Rudaz}]{Kapusta:2020npk}
\bibinfo{author}{Kapusta\xfnm[ J.I.]}, \bibinfo{author}{Rrapaj\xfnm[ E.]},
  \bibinfo{author}{Rudaz\xfnm[ S.]}.
\newblock \bibinfo{title}{{Spin versus Helicity Equilibration Times and
  Lagrangian for Strange Quarks in Rotating Quark-Gluon Plasma}}
  \bibinfo{year}{2020};\href{http://arxiv.org/abs/2004.14807}{\tt
  arXiv:2004.14807}.

\end{thebibliography}

\end{document}